\documentclass{llncs}

\usepackage[T1]{fontenc} 
\usepackage[latin1]{inputenc}
\usepackage{tikz}
\usepackage{amsmath,amssymb}
\usepackage{subfigure}
\usetikzlibrary{arrows,automata,shapes}

\newenvironment{keywords}{
       \list{}{\advance\topsep by0.35cm\relax\small
       \leftmargin=1cm
       \labelwidth=0.35cm
       \listparindent=0.35cm
       \itemindent\listparindent
       \rightmargin\leftmargin}\item[\hskip\labelsep
                                     \bfseries Keywords:]}
     {\endlist}

\begin{document}

\title{Engineering hierarchical complex systems: an agent-based approach}
\subtitle{The case of flexible manufacturing systems}

\titlerunning{Engineering hierarchical complex systems}

\author{Gildas Morvan\inst{1,2} \and Daniel Dupont\inst{1,3} \and Jean-Baptiste Soyez\inst{1,4} \and Rochdi Merzouki\inst{1,4}}
\authorrunning{G. Morvan, A. Veremme, D. Dupont}

\institute{Univ. Lille Nord de France,  1bis rue Georges Lefèvre 59044 Lille cedex, France \and LGI2A, U. Artois, Technoparc Futura 62400 Béthune, France. email: \email{first name.surname@univ-artois.fr}  \and HEI, 13 rue de Toul 59046 Lille Cedex, France. email: \email{first name.surname@hei.fr} \and LAGIS, EC-Lille, Avenue Paul Langevin BP 48 59651 Villeneuve D'ascq cedex, France}

\maketitle

 \begin{abstract}
This article introduces a formal model  to specify, model and validate hierarchical complex systems described at different levels of analysis. It relies on concepts that have been developed in the multi-agent-based simulation (MABS) literature: \textit{level}, \textit{influence} and \textit{reaction}.  One application of such model is the specification of hierarchical complex systems, in which decisional capacities are dynamically adapted at each level with respect to the emergences/constraints paradigm. In the conclusion, we discuss the main perspective of this work: the definition of a generic meta-model for holonic multi-agent systems (HMAS). 
\end{abstract}

\begin{keywords}
multi-level multi-agent based simulations, formal models, hierarchical systems
\end{keywords} 

\section{Introduction}

\setcounter{footnote}{0}

Engineering a complex system such as a flexible manufacturing system (FMS) is a challenging problem. The target system is complex, holonic, relies on distributed decisional processes, and must be adaptive, \textit{i.e.}, robust to perturbations and easily reconfigurable.

To solve these problems, proposed solutions\footnote{\textit{E.g.}, heterarchical~\cite{Duffie:1996} or semi-heterarchical~\cite{Sallez:2010a} control, holonic multi-agent systems (HMAS)~\cite{Bendriss:2007,Gaud:2008,Van-Brussel:1998,VanBrussel:2007} or intelligent product based concepts~\cite{Sallez:2010}.} take advantage of system
\begin{itemize}
\item  complexity, distributing the control in system components that embody primitive cognitive capacities, \textit{e.g.}, be able to be identified, to communicate, to react to environmental changes, 
\item  holonic structure, using dedicated meta-models and conception methodologies.
\end{itemize}

An important tool in the design, simulation and validation of such solutions has been multi-agent-based simulation (MABS). This article introduces a formal model  to specify, model and and validate hierarchical complex systems. It takes inspiration from two trends in MABS research:
\begin{itemize}
\item the formalization of interaction models, 
\item multi-level modeling,  where interacting agents are ontologically distributed among multiple layers of organization. 
\end{itemize}

The article is organized as follows: 
\begin{itemize}
\item in the section~\ref{trends}, the two trends of MABS research cited above, multi-level modeling and formal modeling\footnote{The presentation focuses on the \textit{influences} $\rightarrow$ \textit{reaction} model (IRM). Other approaches such as IODA~\cite{Kubera:2008} or based on DEVS~\cite{Muller:2009} are not described.}, are introduced,
\item  the section~\ref{MLM} presents a generic formal model for multi-level MABS, 
\item an abstract implementation of this model, focusing on the specification of hierarchical multi-agent systems (MAS), in which decisional capacities are dynamically adapted at each level with respect to the emergences/constraints paradigm, is proposed in the section~\ref{emergenceconstraint},
\item the conclusion (section~\ref{conclusion}) summarizes our contributions and perspectives.
\end{itemize}

\section{Two trends in MABS research}
\label{trends}

\subsection{Multi-level modeling}
\label{multi-level}

A \textit{level} represents a point of view on the system, and its relations to other points of view~\cite{Morvan:2011}. While this concept seems important to understand complex systems\footnote{Multi-level approaches have proven useful in many domain such as statistics~\cite{Goldstein:2010}, chemistry~\cite{Horstemeyer:2010,Lucia:2010}, physics~\cite{Steinhauser:2008}, hydrology~\cite{Servat:1998a} or biology~\cite{Uhrmacher:2007}.}, it generally remains abstract: implementations tend to constraint this definition, in particular the relations between levels. Therefore, a \textit{multi-level} model integrates knowledge on different levels and their relations. 
\textit{Multi-scale} are multi-level models characterized by hierarchical relations in levels~\cite{Gil-Quijano:2010,McGregor:2005,Muller:2005,Ratze:2007}. A level may represent, according to the context, a spatio-temporal extent,  a position in a decision hierarchy, etc. Let consider these two examples.
\begin{enumerate}
\item The system is characterized by  processes that have different spatio-temporal extents. Two types of relations can be commonly found  in such models:
\begin{itemize}
\item scaling, \textit{i.e.}, computing macroscopic (resp. microscopic) variables from microscopic (resp. macroscopic) processes,
\item grouping and degrouping (or aggregation and disaggregation)~\cite{Gaud:2008,Navarro:2011,Servat:1998a}, \textit{i.e.}, defining a process at a level as a group (resp. part) of  processes (resp. a process) at an other level.
\end{itemize}
\item Levels are characterized by decisional capacities; relations represent the \textit{emergence} of new capacities and the \textit{constraint} over existing capacities~\cite{Morvan:2008,Morvan:2009a}.
\end{enumerate}


A level is often viewed as a \textit{level of organization}. This concept is closely related to the notion of holon~\cite{Gaud:2008}. This aspect is discussed in the section~\ref{conclusion}.


\subsection{The \textit{influences} $\rightarrow$ \textit{reaction} model}

The \textit{influences} $\rightarrow$ \textit{reaction} model (IRM) has been developed to address issues raised by the classical vision of action in Artificial Intelligence as \textit{the transformation of a global state}~\cite{Ferber:1996}: 
\begin{itemize}
\item simultaneous actions cannot be easily handled,
\item the result of an action depends on the agent that performs it but not on other actions,
\item the autonomy of agents is not respected.
\end{itemize}

 Basically, it decomposes action in two phases:
 agents and environment (micro level) produce a set of influences, then the system (at macro level) reacts to influences; \textit{e.g.}, detects and solves influence conflicts such as in the platform Jaak\footnote{\url{http://www.janus-project.org/Jaak}}.
 As~\cite{Michel:2007} notes, "the influences [produced by an agent] do not directly change the environment, but rather represent the desire of an agent to see it changed in some way". Thus, reaction computes the consequences of agent desires and environment dynamics. In recent years, variants of IRM have been developed to handle specific situations~\cite{Michel:2007,Morvan:2011,Weyns:2003,Weyns:2004}. This presentation focuses on the influence reaction model for simulation (IRM4S)~\cite{Michel:2007}. 
 
Let $\delta(t) \in \Delta$ be the dynamic state of the system at time $t$:
\begin{equation}
   \delta(t) = < \sigma(t), \gamma(t) >, 
\end{equation}
where $\sigma(t) \in \Sigma$ is the set of environmental properties and $\gamma(t) \in \Gamma$ the set of influences, representing system dynamics. The state of an agent $a \in A$ is characterized by its physical state $\phi_a \in \Phi_a$ with $\Phi_a \in \Sigma$ (\textit{e.g.}, its position) and  its internal state $s_a \in S_a$ (\textit{e.g.}, its beliefs).

The evolution of the system from $t$ to $t + dt$ is a two-step process:
\begin{enumerate}
\item agents and environment produce a set of influences\footnote{the sets of producible influence sets and influences produced at $t$ are denoted respectively $\Gamma'$ and $\gamma'(t)$ to point out that the latter is temporary and will be used to compute the dynamic state of the system at $t+dt$.} $\gamma'(t) \in \Gamma'$,
\item the reaction to influences produces the new dynamic state of the system.
\end{enumerate}

An agent $a \in A$  produces influences through a function $Behavior_a: \Delta \mapsto \Gamma'$. This function is decomposed into three functions executed sequentially:
\begin{equation}
	p_a(t) = Perception_a(\delta(t)),
\end{equation}
\begin{equation}
	s_a(t+dt) = Memorization_a(p_a(t), s_a(t)),
\end{equation}
\begin{equation}
	\gamma'_a(t) = Decision_a(s_a(t+dt)).
\end{equation}

The environment produces influences through a function $Natural_\omega: \Delta \mapsto \Gamma'$:
\begin{equation}
	\gamma'_\omega(t) = Natural_\omega(\delta(t)).
\end{equation}

Then the set of influences produced in the system at $t$ is:
\begin{equation}
\gamma'(t) = \{\gamma(t) \cup \gamma'_\omega(t) \cup \bigcup_{a \in A} \gamma'_a(t) \}.
\end{equation}

After influences have been produced, the new dynamic state of the system is computed by a  function $Reaction: \Sigma \times \Gamma' \mapsto \Delta$ such as: 
\begin{equation}
	\label{reaction}
	\delta(t+dt) = Reaction(\sigma(t), \gamma'(t)).
\end{equation}

\section{A generic meta-model for multi-level MABS}
\label{MLM}

In this section, a generic meta-model for multi-level MABS, called IRM4MLS, is presented\footnote{The dynamic aspects of the meta-model, \textit{i.e.}, simulation algorithms, are not described here. An exhaustive presentation can be found in~\cite{Morvan:2011}.}.  This model has the following interesting properties:
\begin{itemize}
\item any valid instance can be simulated~\cite{Soyez:2011},
\item simulation scheduling is logically distributed by level,
\item complexity of simulation algorithm can be optimized according to model structure.
\end{itemize}

\subsection{Specification of the levels and their interactions} 

A multi-level model is defined by a set of levels $L$ and a specification of the relations between levels\footnote{The notion of level is here similar to the notion of \textit{brute space} in the MASQ meta-model~\cite{Stratulat:2009}.}. Two kinds of relations are specified in IRM4MLS: an influence relation (agents in a level $l$ are able to produce influences in a level $l' \neq l$) and a perception relation (agents in a level $l$ are able to perceive the dynamic state of a level $l' \neq l$), represented by directed graphs denoted respectively $<L, E_I>$ and $<L, E_P>$, where $E_{I}$  and $E_{P}$ are two sets of edges, \textit{i.e.}, ordered pairs of elements of $L$. Influence and perception relations in a level are systematic and thus not specified in $E_I$ and $E_P$ (cf. eq.~\ref{I-} and \ref{I+}). 

\textit{E.g.},$\forall l,l' \in L^2$, if  $E_{P} = \{ll'\}$ then the agents of $l$ are able to perceive the dynamic states of  $l$ and  $l'$ while the agents of $l'$ are able to perceive the dynamic state of  $l'$. 


The in and out neighborhood in  $<L, E_I>$ (respectively $<L, E_P>$)  are denoted $N_{I}^-$ and $N_{I}^+$ (resp. $N_{P}^-$ and $N_{P}^+$) and are defined as follows:
\begin{equation}
\forall l \in L, N_{I}^-(l) \mbox{ (resp. } N_{P}^-(l)\mbox{) } =  \{ l \} \cup \{l' \in L: l'l \in E_{I} \mbox{ (resp. } E_{P}\mbox{)} \},
\label{I-} 
\end{equation}
\begin{equation}
\forall l \in L, N_{I}^+(l) \mbox{ (resp. } N_{P}^-(l)\mbox{) } =  \{ l \} \cup \{l' \in L: ll' \in E_{I} \mbox{ (resp. } E_{P}\mbox{)} \},
\label{I+} 
\end{equation}

\textit{E.g.}, $\forall l,l' \in L^2$ if $l' \in N_I^+(l)$ then the environment and the agents of $l$ are able to produce influences in the level $l'$; conversely we have $l \in N_I^-(l')$, \textit{i.e.}, $l'$ is influenced by $l$.

 \subsection{Agent population and environments} 

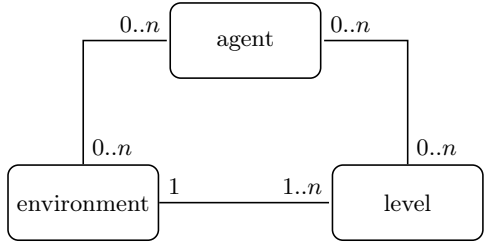
\begin{figure}[t]
	\begin{center}
		\tikzstyle{blocdebase}= [draw, text centered,minimum height=1cm, minimum width=2cm, rounded corners]

		\begin{tikzpicture}[-,>=stealth',shorten >=1pt,auto,semithick]
   			\node[blocdebase] (agent)  {agent};
   			\node[blocdebase,node distance=\textwidth/4] (environnement)[below left 
of=agent] {environment};
			\node[blocdebase,node distance=\textwidth/4] (niveau)[below right 
of=agent] {level};
    			    
    			\draw (environnement.north) |- (agent.west) ;
			\draw  (niveau.north) |- (agent.east);
			\draw (environnement) -- (niveau);
			
			\node [above right] at (niveau.north) {$0..n$}; 
			\node [above left] at (niveau.west) {$1..n$}; 
			\node [above left] at (agent.west) {$0..n$}; 
			\node [above right] at (agent.east) {$0..n$}; 
			\node [above right] at (environnement.north) {$0..n$}; 
			\node [above right] at (environnement.east) {$1$}; 
						
		\end{tikzpicture}
		\caption{Main concepts of IRM4MLS (cardinalities are specified in the UML fashion)}
		\label{conceptIRM4MLS}
	\end{center}
\end{figure}

The set of agents in the system at time $t$ is denoted $A(t)$. $\forall l \in L$, the set of agents belonging to $l$ at $t$ is denoted $A_l(t) \subseteq A(t)$. An agent belongs to a level iff a subset of its physical state $\phi_a$ belongs to the state of the level:
 \begin{equation}
\label{indicatorfunction}
\forall a \in A(t), \forall l \in L, a \in A_l(t) \text{~iff~} \exists \phi^l_a(t) \subseteq \phi_a(t) | \phi^l_a(t) \subseteq \sigma^l(t).
\end{equation}
Thus, an agent belongs to zero, one, or more levels. As notes~\cite[p. 815]{Stratulat:2009}, the physical state of an agent in a level, \textit{i.e.}, its \textit{body},  is "the manifestation of an agent in the environment and allows others to perceive it."
An environment can also belong to multiple levels (cf. fig.~\ref{conceptIRM4MLS}).

 \subsection{Action modeling} 
 
 The dynamic state of a level $l \in L$ at time $t$, denoted  $\delta^l(t) \in \Delta^l$, is a tuple  $< \sigma^l(t), \gamma^l(t) >$, where $\sigma^l(t)  \in \Sigma^l$ and  $ \gamma^l(t) \in  \Gamma^l$ are the sets of environmental properties and influences of $l$. 
 
The influence production step takes into account the influence and perception relations between levels:
\begin{equation}
\forall a \in A_l, Behavior_a^l:  \prod_{l_P \in N_P^+(l)}  \Delta^{l_P}  \mapsto \prod_{l_I \in N_I^+(l)} \Gamma^{l_I}{'}.
\end{equation}

Once influences have been produced, interactions between levels do not matter anymore. Thus,  the reaction function defined in IRM4S can be re-used: 
 \begin{equation}
	Reaction^l: \Sigma^l \times \Gamma^l{}' \mapsto \Delta^l,
\end{equation}
where $Reaction^l$ is the reaction function proper to each level. 

\section{Engineering hierarchical complex systems with IRM4MLS}

\label{emergenceconstraint}

\subsection{The emergence/constraint paradigm}
In many MABS, processes are considered on the following 2-level  relative hierarchy:
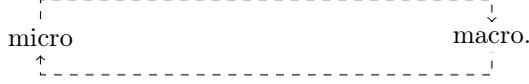
\begin{center}
\begin{tikzpicture}
\node (a) at (0,0) {micro} ;
\node (b) at (6,0) {macro.} ;
\draw [->,dashed]  (a)  -- ++(0,.5)  -|  (b); 
\draw [<-,dashed]  (a)  -- ++(0,-.5)  -|  (b); 
\end{tikzpicture} 
\end{center}

Arrows represent causality relations between levels. Dashing suggests that they are generally not explicitly defined but emerge from interactions between entities. \textit{A contrario}, a multi-level approach considers these relations explicitly. In engineering applications, a level may rather represents a position in a decision hierarchy (cf. section~\ref{multi-level}). Two kinds of relation may be distinguished in such systems: \textit{emergence} of new capacities and \textit{constraint} over existing capacities~\cite{Morin:1992}.  Let consider an example in the domain of FMS engineering. In a case study on automated guided vehicle (AGV) control presented in~\cite{Morvan:2009a} (cf. section~\ref{casestudy}), the model relies on the following relations:
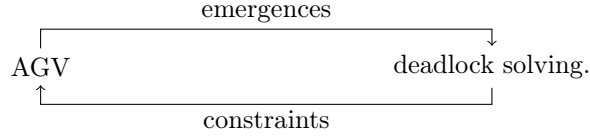
\begin{center}
\begin{tikzpicture}
\node (a) at (0,0) {AGV} ;
\node (b) at (6,0) {deadlock solving.} ;
\node at (3,.7)  {emergences} ;
\node at (3,-.7)  {constraints} ;
\draw [->]  (a)  -- ++(0,.5)  -|  (b); 
\draw [<-]  (a)  -- ++(0,-.5)  -|  (b); 
\end{tikzpicture} 
\end{center}
Macro agents (representing a set of "trapped" AGVs) emerge from micro agent interactions when an interaction pattern defined as a deadlock is detected, and then constraint
their behaviors to solve it. While the notions of emergence and constraint were informally defined in~\cite{Morvan:2009a}, formal definitions in the context of IRM4MLS are given in the following.

\subsection{IRM4MLS implementation}

Let $L$ be a hierarchy and $\{\mu, M\} \subseteq L$ two hierarchically coupled levels, $\mu$ referring to the micro level and $M$ to the macro level. Thus, $A_\mu$ (respectively $A_M$) denotes the agents of the micro-level  (resp. macro-level).  The  emergence/constraint paradigm supposes that $E_I \supseteq \{\mu{}M, M\mu\}$. 

\begin{equation}
\forall l \in L, \gamma^l{'}(t) = \{\gamma^l(t), \gamma^{M}_{\omega}, \gamma^{\mu}_{\omega},  \bigcup_{a \in A_M} \gamma_a^{M}{'}(t), \bigcup_{a \in A_\mu} \gamma_a^{\mu}{'}(t)  \}.
\end{equation}

An \textit{emergence} $e$ at the level $M$ is an influence that has the following properties: 
\begin{itemize}
\item $e$ belongs to the macro-level but not to the micro-level: 
\begin{equation}
e \in \Gamma^{M}  \text{~but~} e \notin \Gamma^{\mu},
\end{equation}
\item $e$ cannot be produced by the behavior of an agent or the environment of $M$: 
\begin{equation}
\forall t, e \notin \bigcup_{a \in A_M}  Behavior_a^M(\delta(t)) \cup Natural_\omega^M(\delta(t)), 
\end{equation}
with $\delta(t)=< \delta^M(t), \delta^\mu(t) >$.
\end{itemize}
Emergent influences generally determine the life-cycle (creation, evolution, destruction) of agents at the macro-level. 

A \textit{constraint} over an influence $i$, denoted $\neg i$, is the special kind of influence that has the following properties: 
\begin{itemize}
\item $\{i, \neg i \}$ belongs to the micro-level but not to the macro-level: 
\begin{equation}
\{i, \neg i \} \subseteq \Gamma^{\mu}   \text{~but~} \{i, \neg i \} \nsubseteq \Gamma^{M},
\end{equation}
\item $\neg i$ cannot be produced by the behavior of an agent or the environment of $\mu$: 
\begin{equation}
\forall t, \neg i \notin \bigcup_{a \in A_\mu} Behavior_a^\mu(\delta(t)) \cup Natural_\omega^\mu(\delta(t)),
\end{equation}
with $\delta(t)=< \delta^M(t), \delta^\mu(t) >$,
\item $\neg i$ inhibits $i$:
\begin{equation}
\begin{array}{c}
 \text{if~} \{i, \neg i\} \subseteq \gamma^\mu{}'(t) \text{~then~}\\ Reaction^\mu(\sigma^\mu{}(t), \gamma^\mu{}'(t)) = Reaction^\mu(\sigma^\mu{}(t), \gamma^\mu{}'(t) \backslash \{i\}).  
\end{array}
\end{equation}
\end{itemize}


\subsection{Conception of hierarchical systems}

The approach described below can be viewed as a semi-heterarchical  control one and takes advantage of complexity and hierarchical (not yet holarchical) organization of the system, distributing the control by level. Heterarchical control methods rely on self-organization principles\footnote{Self-organized systems are generally characterized by the use of environment as a communication medium to carry local informations as well as positive and negative feedbacks.}  and therefore assume that the system is able to achieve its goals and is easily reconfigurable, \textit{i.e.}, that the normal functioning mode emerges from the interactions between system components (products, machines, simulated entities, etc.) that embody limited cognitive capabilities (cf. introduction). However, the trajectory of such systems may lead to non desired attractors.

\begin{figure}[h]
\begin{center}
\begin{tikzpicture}[node distance=3cm]
\node (ss) {Specification} ;
\node (sd) [right of=ss] {Model design} ;
\node (cd) [right of=sd]  {Control design} ;
\node (va) [right of=cd] {Implementation} ;

\path	
  (ss)  edge [->] node{} (sd) 
  (sd)  edge [->] node[above] {} (cd) 
  (sd)  edge [->, loop below] node[below]{\textit{verification}} (sd)
  (cd)  edge [->] node[above] {} (va)
  (cd)  edge [->, loop below] node{\textit{validation}} (cd)
  (cd)  edge [->, bend right] node[above] {\textit{modification}} (ss);
  
%
%
\path[rounded corners,draw] (sd.south west) rectangle (cd.north east);

\end{tikzpicture}
 \caption{Engineering methodology}
  \label{methodo}
 \end{center}
\end{figure}
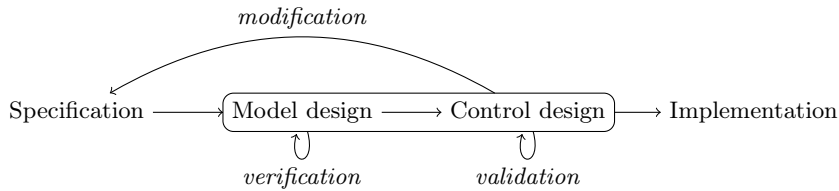

The proposed methodology is presented in the fig.~\ref{methodo}. The system is designed iteratively in a two-step process.
\begin{enumerate}
\item From an initial specification of the system, a model of the system in normal functioning mode, is defined and verified, \textit{i.e.}, that system components have the necessary cognitive capacities to perform their tasks.
\item From non desired attractors exhibited by the simulation of the model, the control strategy may be designed and validated. However, it is likely that the specification of the system has be modified to do so, \textit{e.g.}, because a new decisional level is needed.
\end{enumerate}

The notion of influence is very general and therefore, may have many possible meanings. In this case,  let
\begin{itemize}
\item  $\gamma^l{'}(t)$ be the capacities of each agent of a level $l$ at time $t$, \textit{i.e.}, the tasks they can perform at the moment, 
\item $\gamma^{l}(t)$ the actual affectation of tasks to agents; the only cognitive capacity required for agents is to expose services they may provide.
\end{itemize}
Thus, $Reaction^l$ is a task assignment algorithm that computes $< \sigma^l(t),  \gamma^l(t) >$ from $< \sigma^l(t), \gamma^l{'}(t) >$.
Note that the hierarchical nature of the system allows to decompose the specification of the system $S$ by level: 
\begin{equation}
S = \{\gamma^l(\delta^l) : \forall l \in L, \forall \delta^l \in \Delta^l \},
\end{equation}
\textit{i.e.}, task assignments for all functioning modes.

That design should lead to the definition of reaction functions that \textit{control} goal affectations. If such a function cannot be defined, then the system design is not valid and must be redefined. This process is iterated until a solution is found (cf. fig.~\ref{methodo}).

\subsection{Case study: AGV deadlocks in gradient field-based FMS} 
\label{casestudy}
The main functionalities of an intelligent transportation system (ITS) are: (1) transport assignment, (2) routing, (3) gathering traffic information, (4) collision avoidance, (5) deadlock avoidance~\cite{Weyns:2008}.

Gradient field-based approaches, where AGV trajectories are computed from gradient fields, allow to implement efficient ITS in FMS~\cite{Ueda:2001,Ueda:1997}. A dedicated task assignment algorithm is generally used to ensure functionality 1, while functionalities 2--4 rely on AGV and shop self-organization properties. Thus, an AGV has two cognitives capabilities: sense attractive or repulsive force fields and emit a repulsive force field.  Similarly, a shop is able to emit attractive fields to require products to process and give back the result to the system. A known problem of  gradient field-based approaches is that a group of AGVs may be trapped in local minima that lead to a system deadlock~\cite{Ueda:2006,Ueda:1997,Weyns:2006}. However, this issue can be easily addressed by hierarchical control methods that compute explicit trajectories\footnote{Readers interested in general, \textit{i.e.}, not gradient-field based approaches, deadlock avoidance techniques in FMS may refer \textit{e.g.}, to~\cite{Banaszak:1990,Ezpeleta:2002,Yoo:2005}.}. 

\begin{figure}[t]
\begin{center}
\subfigure[Initial design]{
\begin{tikzpicture}[node distance=2.2cm]
\node (system) {Task assignment} ;
\node (agv) [below left of=system] {AGV} ;
\node (shop) [below right of=system] {Shop} ;
\path
  (system)  edge [->,right] node{$\neg i$} (agv) 
  (system)  edge [->, bend left] node[right] {$\neg i$}  (shop)
  (shop)  edge [->,bend left] node[left] {$e$}  (system) 
  (agv)  edge [<->,dashed] node{}  (shop) ; 
\end{tikzpicture} 
\label{example:subfig1}
}
\subfigure[Final design]{
\begin{tikzpicture}[node distance=2.2cm]
\node (system) {Task assignment} ;
\node (agv) [below left of=system] {AGV} ;
\node (congestion) [above left of=agv]  {Deadlock solving} ;
\node (shop) [below right of=system] {Shop} ;
\path
  (system)  edge [->,right] node{$\neg i$} (agv) 
  (system)  edge [->, bend left] node[right] {$\neg i$}  (shop)
  (shop)  edge [->,bend left] node[left] {$e$}  (system) 
  (congestion)  edge [->, bend left] node[right] {$\neg i$}  (agv)
  (agv)  edge [->,bend left] node[left] {$e$}  (congestion) 
  (agv)  edge [<->,dashed] node{}  (shop) ;
  (system)  edge [<->,dashed] node{}  (congestion) ; 
\end{tikzpicture} 
\label{example:subfig2}
}
\begin{tikzpicture}[node distance=4cm]
\node (a) {};
\node (b) [right of=a]  {mediated interactions}; 
\path (a.east)  edge [<->,dashed] node{} (b.west) ;
\end{tikzpicture} 
\caption{Decisional levels in the case study on AGV control}
\label{example}
\end{center}
\end{figure}
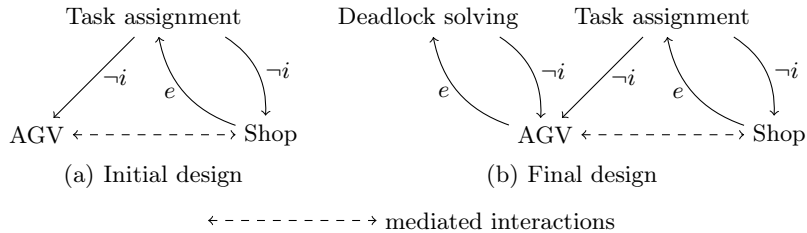

The first design of the system is presented in fig.~\ref{example:subfig1}:  a task assignment algorithm affects goals to AGVs (statically, a signal to maximize) and shops (dinamically, products to process).  The deadlock avoidance functionality is not explicitly programmed but is supposed to emerge from mediated interactions between AGVs and shops. Various researches have shown that such a solution may reduce the number of deadlock occurrences but not eliminate it: routing is not deadlock avoidance~\cite{Weyns:2006}. A new system architecture is then designed (cf. fig.~\ref{example:subfig2}): if a deadlock (reified by an emergence) is  observed by a deadlock solving algorithm, constraints over signal sensing and emission are computed to solve it~\footnote{Practical aspects of this approach are discussed in~\cite{Morvan:2009a}. \textit{E.g.}, AGVs embody the deadlock solving algorithm, becoming multi-level agents. This problem has been an important motivation in the development of IRM4MLS.}. 


\section{Conclusion}
\label{conclusion}

In this article, we have presented a formal model for MABS and its implementation to engineer hierarchical complex systems. Two types of influences have been distinguished in this approach : \textit{emergence}, that basically triggers a new system behavior when a specific pattern is detected (in the  previous short example of  gradient field-based FMS, the detection of a deadlock triggers the modification of AGV repulsive signal emission) and \textit{constraint}, that, as its name suggests, constraints decisional capacities of system entities to solve a situation. 

The main advantage of this approach lies in the multi-level and simulation capabilities of IRM4MLS, to model a system in which decisional capacities are distributed in its components and evolve along time to meet user's goals and to simulate a model whiteout bias and temporal deadlocks\footnote{Simulation properties of IRM4MLS may be exploited to explore model behavior using, \textit{e.g.}, the polyagent concept~\cite{Parunak:2007,Parunak:2011}. Such an approach may be used to determine fail probabilities of system components or control strategies.}.  Its main drawback is the strict hierarchical organization in levels. 

Holonic multi-agent systems (HMAS) can be viewed as a specific case of multi-level multi-agent-systems (MAS), the most obvious aspect being the \textit{loosely hierarchical} organization of levels. However, from a methodological perspective, differences remain: thus, most of holonic meta-models  focus on organizational aspects (cf. \textit{e.g.},~\cite{Bendriss:2007,Gaud:2008,Van-Brussel:1998,VanBrussel:2007}). An important issue towards a generic meta-model for HMAS would be to define a holon with respect to IRM4MLS concepts: a \textit{holon} cannot be defined with IRM4MLS first class abstractions (level, agent or environment), as it represents a multi-level entity. This situation is the main perspective of this work.

\section*{Acknowledgments}

Authors would like to thank Daniel Jolly (LGI2A, Université d'Artois, Béthune France) and Alexandre Veremme (HEI, pôle recherche Ingénierie et Sciences du Vivant, Lille France) for their help and support. Jean-Baptiste Soyez is funded by the InTrade project\footnote{\url{http://www.intrade-nwe.eu}}. 

\bibliographystyle{splncs03.bst}
\bibliography{../../Biblio}

\end{document}